# Persistence of large and gate-tunable anisotropic magnetoresistance in an atomically thin antiferromagnet

Cheol-Yeon Cheon[1, 2 *], Kenji Watanabe[3], Takashi Taniguchi[4], Alberto F. Morpurgo[1, 2 *], and Dmitry Lebedev[1, 2 *]

1. Department of Quantum Matter Physics, University of Geneva, 24 Quai Ernest Ansermet, CH-1211 Geneva, Switzerland
2. Department of Applied Physics, University of Geneva, 24 Quai Ernest Ansermet, Geneva, CH-1211 Switzerland
3. Research Center for Electronic and Optical Materials, National Institute for Materials Science, 1-1 Namiki, Tsukuba 305-0044, Japan
4. Research Center for Materials Nanoarchitectonics, National Institute for Materials Science, 1-1 Namiki, Tsukuba 305-0044, Japan

*Correspondence should be addressed to: cheolyeon.cheon@unige.ch, alberto.morpurgo@unige.ch, and dmitry.lebedev@unige.ch

**ABSTRACT**

Anisotropic magnetoresistance (AMR) offers a robust electrical readout of antiferromagnetic (AFM) states, playing a central role in the rapidly advancing field of AFM spintronics. Despite its great versatility, electrical probing of the Néel vector via AMR remains challenging in the ultrathin limit due to interface disorder and reduced dimensionality. Here, we demonstrate electrical readout of the Néel vector down to 1.3 nm (two layers) in the two-dimensional van der Waals (vdW) AFM semiconductor $NiPS_3$. Leveraging spin-flop-mediated rotation of the Néel vector and using both transistor and tunnel-junction device geometries, we identify two distinct AMR contributions in $NiPS_3$, that dominate at low and high charge densities, respectively. We achieve full gate control over these contributions, enabling tunability of both the magnitude and sign of magnetoresistance. Our results establish semiconducting vdW antiferromagnets as a rich platform for studying AMR in the ultrathin limit, opening new avenues for multifunctional AFM spintronic devices.

Layered two-dimensional (2D) van der Waals (vdW) magnets have emerged as a promising platform for spintronics research.[1] Their atomic-scale thickness enables novel functionalities, including the ability to tune spintronic behavior through the application of external electric fields and heterostructure engineering.[2] The 2D nature of vdW magnets offers unique opportunities to probe and exploit the interplay between electrical transport and layer-number dependent magnetic order.[3,4] Among these materials, transition-metal phosphorus chalcogenides (e.g., $NiPS_3$, $CrPS_4$) have recently attracted attention.[5,6] These antiferromagnetic (AFM) semiconductors exhibit substantially higher in-plane carrier mobility well below the magnetic transition temperature than the archetypal chromium trihalides, enabling electrical access to diverse magnetic configurations that persist down to the monolayer limit.[7,8] Their semiconducting character can further unlock highly desirable functionalities, such as gate-tunability and coupling to light – capabilities that remain largely unexplored as majority studies in AFM spintronics have so far focused on metals.[9,10]

In AFM spintronics, where information is encoded in the staggered magnetization (or simply Néel vector for bipartite magnet), manipulating the spin configuration is achieved through spin-torque-driven mechanisms, while the electrical readout is commonly based on anisotropic magnetoresistance (AMR).[11–16] The spin-torque-driven switching generally benefits from scaling down due to reduced number of spins to manipulate. In contrast, AMR – the dependence of a material's resistivity on the orientation of its magnetic order relative to the electrical current and/or crystal axes – is significantly suppressed in the ultrathin limit. In particular, enhanced surface and interface scattering complicate the electrical readout[17–19], while reduced dimensionality destabilizes long-range AFM order.[20] As a result, robust AMR has so far been reliably demonstrated only in films with thicknesses on the order of tens of nanometers[9,21–23], and it remains an open question whether the Néel vector in ultrathin AFM

materials – approaching atomically thin limit – can be directly and reliably probed by purely electrical means.

In this work, we establish AMR as a reliable probe of the Néel vector in the vdW AFM semiconductor NiPS$_3$. We detect the AMR signal from the in-plane Néel vector rotation in NiPS$_3$ by leveraging the spin-flop transition with in-plane magnetic field. We demonstrate robust AMR signal down to 1.3 nm (two layers), representing – to the best of our knowledge – the thinnest channel in which AMR signal from AFM material has been detected. Furthermore, we show that the magnitude and sign of the magnetoresistance can be tuned electrostatically using gate, revealing two distinct contributions in AMR mechanism that depend on carrier doping in NiPS$_3$. Our results show that an electrically tunable readout of the Néel vector in the ultrathin limit can be achieved with vdW AFM semiconductors, a functionality that is essential for the development of electrically tunable 2D AFM spintronic devices.

NiPS$_3$ is a layered semiconductor with collinear AFM order below the Néel temperature of 140-150 K in multilayer crystals (**Figure 1a**).[5,24] The spins almost entirely lie in the layer plane and are aligned along the crystallographic *a*-axis (i.e. magnetic easy axis, **Figure 1b**). To investigate AMR in NiPS$_3$, we fabricated devices with two orthogonal directions of electrical current: transistors with in-plane electrical current and tunnel junctions with the current out-of-plane. (**Figure 1c**). All the devices were assembled via dry pick-up and transfer methods using multilayer NiPS$_3$ as a channel material, graphite flakes as contacts, and h-BN crystals for encapsulation (see Methods). The transport characteristics of these devices confirm their high quality and show the expected behavior: n-type semiconducting switching for field-effect transistor (FET) devices and non-linear IV characteristics for vertical tunnel junctions (**Figure 1d, e**).[5,25]

To gain control over orientation of Néel vector, which is needed to study AMR in a collinear antiferromagnet,[26] we exploit the spin-flop transition of NiPS$_3$.[5,27] Specifically, when an in-plane magnetic field is applied along the easy axis of NiPS$_3$, the spins rotate and become oriented perpendicular to the easy axis above the spin-flop field of $B_{sf} \approx 10.5$ T, while remaining in the layer plane because of the strong easy-plane anisotropy (**Supplementary Figure S1**).[27,28] Indeed, recent optical studies on bulk NiPS$_3$ directly captured this field-driven Néel vector rotation,[27,28] and showed that its in-plane angular position is accurately described within mean-field theory by:[29]

$$\tan 2\beta(B) = \frac{\sin 2\beta(0\text{T})}{\cos 2\beta(0\text{T}) - \frac{B^2}{B_{sf}^2}} , \qquad (1)$$

where $\beta(B)$ is the azimuthal angle between magnetic field and Néel vector at a given magnetic field strength, $\beta(0\text{T})$ is the angle between magnetic field and easy axis. Owing to the layered nature of NiPS$_3$ and persistence of the uniaxial magnetic anisotropy down to bilayer thickness,[5] spin-flop–mediated control of the Néel vector provides a versatile approach for ultrathin samples. Magnetic order of monolayer NiPS$_3$ is governed by hexagonal magnetic anisotropy[5] and therefore cannot be accurately described by Eq. (1), hence in the present work we investigate AMR down to bilayer thickness.[30]

With the Néel vector orientation in NiPS$_3$ controlled by the magnetic field, we analyze the angular dependence of the resistance using a commonly employed expression for AMR that includes the so-called noncrystalline and crystalline contributions.[26,31] The former depends on the angle between Néel vector and electrical current, whereas the latter depends on the angle between Néel vector and the crystallographic axes. The crystalline contribution is expected to be sizable in our NiPS$_3$ devices because the FET channel is formed of micron-scale single crystal NiPS$_3$. Because of the monoclinic symmetry of multilayer NiPS$_3$ resulting in uniaxial magnetic anisotropy in bilayer and thicker layers,[5,32] we consider the leading two-fold contributions to the longitudinal resistivity ($\rho_{xx}$):

$$\frac{\rho_{xx}}{\rho_{avg}} = 1 + C_I * \cos 2\alpha(B) + C_U * \cos 2\psi(B) \qquad (2)$$

where $\rho_{avg}$ is resistivity averaged over in-plane Néel vector rotation, $C_I$ and $C_U$ denote noncrystalline and crystalline AMR coefficients, respectively, $\alpha(B)$ is the angle between the Néel vector and electrical current, and $\psi(B)$ is the angle between the Néel vector and the crystallographic $a$-axis (easy axis) at a given magnetic field strength (the angles $\alpha(B)$ and $\psi(B)$ are readily expressed in terms of $\beta(B)$ using a known geometry of the device, see **Supplementary Note 1**).

We begin by measuring magnetoresistance ($MR(B) = \frac{R(B)-R(0)}{R(0)} * 100$ (%)) of 4-nm (6-layer) NiPS$_3$ FET device with in-plane magnetic field oriented near the easy axis (**Figure 2a**). The easy-axis ($a$-axis) direction of the device was determined optically from the direction of linear polarization of its photoluminescence.[5] In order to maximize the angular range of $\beta(B)$, we intentionally align the magnetic field at small positive or small negative angle with respect to the easy axis, i.e., $\beta(0T) \approx \pm 3°$. This enables us to induce a clockwise or counterclockwise rotation of a Néel vector, covering the combined rotation range of almost 180° for $\beta(B)$ as the magnitude of the field is swept from 0 T to ±12 T. We find that MR curves for both field alignments (**Figure 2c**) show a sharp change near a critical field of 10.2 T due to spin-flop, however opposite directions of the Néel vector rotation result in a different sign of MR: a peak for $\beta(0T) \approx + 3°$ and a dip for $\beta(0T) \approx - 3°$.

Note that the direction of electrical current in the 4-nm device is set by the graphite layers and is not parallel to the $a$-axis (**Figures 2a** and **2b**). As a result, clockwise and counterclockwise rotations of the Néel vector change the angle between the Néel vector and the direction of electrical current ($\alpha(B)$) in opposite ways: clockwise rotation decreases the $\alpha(B)$, whereas counterclockwise rotation increases the $\alpha(B)$ (see the insets of **Figure 2c**). We observe that these rotations produce opposite signs of the MR, which indicates that the dominant contribution in the AMR is the $\alpha(B)$-dependent term, i.e., noncrystalline term.[26] Indeed, fitting the AMR with Eq. (2) shows that the noncrystalline AMR amplitude ($2C_I$) is dominant for both clockwise and counterclockwise Néel vector rotations: $2C_I = 0.45 \pm 0.01$ and $2C_U = 0.02 \pm 0.01$ for $\beta(0T) \approx +3°$; $2C_I = 0.38 \pm 0.01$ and $2C_U = 0.03 \pm 0.02$ for $\beta(0T) \approx -3°$ (see **Supplementary Note 2** and **Figure S2** for fitting details). When plotted as a function of $\alpha(B)$ (**Figure 2d**), the resistance data reveal a positive noncrystalline AMR in NiPS$_3$, corresponding to higher (lower) resistivity when electrical current is parallel (perpendicular) to the direction of Néel vector. This dominate noncrystalline AMR can further explain the MR curves measured at much larger $\beta(0T)$ of $\pm 30°$, i.e. when magnetic field is significantly away from the easy axis (see **Supplementary Figure S3)**.

The data discussed above for 4-nm NiPS$_3$ FET device were obtained at a relatively large gate-voltage offset $\Delta V_{BG} = V_{BG} - V_{TH} = 40$ V (corresponding to an electron density of approximately $3 \times 10^{12}$ cm$^{-2}$; $V_{BG}$ and $V_{TH}$ are the back-gate and threshold voltages, respectively). Measuring the same device at lower $\Delta V_{BG}$, we find that the peaks in *MR* around $B_{sf}$ disappear as $V_{BG}$ approaches $V_{TH}$, which corresponds to a sign change of the *MR* at $B = B_{sf}$ from positive to negative (**Figure 3a-b** and **Supplementary Figure S4**). Importantly, at $\Delta V_{BG} \approx 0\,V$, the MR curves for $\beta(0T) \approx \pm 3°$ overlap (**Figure 3c**), indicating that the noncrystalline MR contribution has vanished. These results indicate that at low charge densities ($\Delta V_{BG} \approx 0V$) the resistance exhibits a qualitatively different behavior from that at high $\Delta V_{BG}$, such that the AMR response is governed by the direction of the Néel vector relative to the easy axis (crystalline AMR) and is independent on the direction of the electrical current (vanishing noncrystalline AMR). Indeed, when we plot the resistance data at $\Delta V_{BG} \approx 0\ V$ as a function of angle between the Néel vector and the easy axis ($\psi_B$), the data for the two opposite Néel vector rotations follow a sinusoidal curve (**Figure 3e**).

We repeat the analysis of the MR measured at different gate voltage and use Eq.(2) to extract the noncrystalline ($C_I$) and crystalline ($C_U$) coefficients (**Figure 3d**) as a function of $\Delta V_{BG}$. This procedure enables a quantitative determination of how the two contributions evolve with carrier density. The fitting reveals a crossover of the dominant resistance contribution from positive noncrystalline AMR at high charge density (comparable or higher than $3 \times 10^{12}$ cm$^{-2}$) to positive crystalline AMR at low charge density near the threshold voltage. Note that this separation of the AMR contributions is possible primarily because, in the 4-nm device geometry, the electric current is not aligned with the easy axis. When the electrical current is parallel to the easy axis, the angular dependences of the noncrystalline and crystalline AMR in Eq. (2) become identical and should produce the same negative MR curve at both low and high gate voltage, which we have verified using additional device (**Supplementary Figure S5**).

We confirm our observations of crystalline and noncrystalline AMR by studying NiPS$_3$ tunnel junction devices (**Figure 4a**). In this geometry, the angle between electrical current and Néel vector remains unchanged during the spin-flop transition, which eliminates the noncrystalline AMR contribution and reveals purely the crystalline AMR.[33] We study tunneling devices with NiPS$_3$ barrier thickness ranging from 3 to 22 nm and find nonlinear $IV$ response at T=10 K (**Figure 4b**). Tunneling at high bias occurs in the Fowler−Nordheim regime as indicated by the linear dependence of slope of $ln\frac{J}{V^2}$ vs $\frac{1}{V}$ on the barrier thickness (**Figure 4c**).[34] In this regime, we measured tunneling MR for the thinnest (3 nm) and thickest (22 nm) barriers under in-plane magnetic field oriented along the easy axis (**Figure 4d**). For both thicknesses, the MR curves exhibit a monotonic decrease above $B_{sf}$, which is identical to the behavior observed in lateral NiPS$_3$ FETs at $\Delta V_{BG} \approx 0V$. This observation confirms the presence of crystalline AMR in NiPS$_3$ at low charge densities. The higher percentage of MR in vertical tunneling geometry compared to lateral devices likely originates from even lower charge density in the tunneling regime, at which the lateral FET conduction is unfeasible due to very large channel resistance.

Finally, the vdW nature of NiPS$_3$ enables us to push the device thickness down to few-layer and thereby directly probe the scaling limit of the AMR effect – an issue of central importance for the future miniaturization of AFM-based spintronic devices, including memory elements. We find that the gate-dependent AMR is consistently reproduced across all measured multilayer NiPS$_3$ FET devices down to bilayer thickness (**Figure 5** and **Supplementary Figures S6 and S7**). Remarkably, bilayer NiPS$_3$ with channel thickness of 1.3 nm exhibits identical gate-tunable modulation of both amplitude and sign of the effect (**Figure 5b, c**) as to 4-nm-thick channel discussed above. This indicates that the same crystalline AMR governs the MR response for all the thickness down to bilayer at low charge density regime, which is supported by the monoclinic two-fold symmetry of multilayer NiPS$_3$ induced by its layer stacking[5,32]. When comparing devices spanning different thicknesses, we observe that in the lateral FET geometry the AMR magnitude remains effectively thickness-independent throughout the entire thickness range (**Figure 5d**). As mentioned earlier, magnetism in monolayer NiPS$_3$ is governed by hexagonal magnetic anisotropy[5], which makes the spin-flop-mediated control of the magnetic order described by Eq. (1) valid down to the bilayer limit.

Our experiments establish AMR in vdW AFM semiconductor NiPS$_3$, allowing us to electrically read out the magnetic order parameter (Néel vector) through the spin-flop transition. We further demonstrate that the AMR amplitude persists unchanged down to bilayer thickness of 1.3 nm, and reveal the gate tunability of the AMR due to the competition of the crystalline and noncrystalline contributions, which enables the sign and magnitude of MR to be tuned in the atomically thin regime.

We suggest a possible simple scenario to explain the gate-dependent AMR behavior for both lateral and vertical device geometries. For large $\Delta V_{BG}$ the AMR response is governed by the relative orientation between the Néel vector and the direction of the electrical current. The positive sign of AMR in this regime implies that the resistivity is larger when the electrical current is along the Néel vector, i.e. along the ferromagnetically aligned spins forming the zig-zag chains. This behavior is qualitatively similar to what happens in ferromagnetic systems, where positive AMR is commonly observed and associated with anisotropic scattering of the charge carriers.[26] In contrast, at low doping densities – near the transistor threshold of FETs and in vertical tunnel junctions – the AMR response is governed by an angle between the Néel vector and the crystallographic easy axis. We propose that – owing to spin-orbit interactions – rotation of the Néel vector from the *a*-axis toward the *b*-axis (spin-flop) results in lowering of the conduction band minimum in $NiPS_3$, as predicted earlier in other 2D vdW AMF materials.[35] For tunneling devices the shift of the conduction band reduces the effective barrier height and enhances the electron transmission, while in lateral $NiPS_3$ FET devices it increases the thermally activated carrier density at a given gate voltage, thereby enhancing channel conductance, in both cases leading to a negative magnetoresistance.

Finally, and particularly important in the context of AFM spintronics, our results show that AMR can be pushed down to atomically thin limit, extending beyond what has been achieved in conventional non-vdW AFM materials. To elaborate on this point further, we benchmark our AMR response against other AFM materials (**Figure 5d, e**) and show that $NiPS_3$ is both the thinnest functioning AMR channel to date (1.3 nm, lateral FET) and has the highest AMR magnitudes reported for AFM systems (in vertical tunnel junctions) without observable signal degradation upon thinning. Indeed, in the broader AFM literature, materials such as CuMnAs and Mn$_2$Au show AMR magnitude ($\lesssim$0.1–1%) comparable to what is observed in NiPS$_3$, but their signal unavoidably decreases as the film thickness is reduced, likely due to enhanced interfacial disorder and increased sensitivity to structural imperfections. In contrast, NiPS$_3$ has a vdW layered structure with robust antiferromagnetism and together with the high transport quality of our FET devices allows AMR-based readout of AFM order in the ultrathin limit. More importantly, the semiconducting nature of NiPS$_3$ enables complete tuning between the two AMR regimes, owing to the electrostatic gate control of magnetic and electronic coupling not achievable in metallic systems. On a fundamental level, this allows isolating and investigating each contribution independently, deepening our understanding of the underlying mechanisms. On a practical level, the full electrical controllability of the AMR signal achieved in ultrathin vdW AFM semiconductors highlights their promise for multifunctional device applications in AFM spintronics.

**METHODS**

**Device fabrication**

Bulk NiPS$_3$ crystal was obtained from a commercial source (HQ graphene, grown by CVT-method). Few-nm thick NiPS$_3$ crystals were mechanically exfoliated onto a SiO$_2$/Si substrate inside a nitrogen-filled glovebox maintained at sub-ppm oxygen and water level. The heterostructures of NiPS$_3$, h-BN, and few-layer graphite were assembled inside the same glovebox using the vdW pick-up transfer method with a polycarbonate film. The thickness of NiPS$_3$ was determined by its optical contrast and atomic force microscopy. Electron-beam lithography, reactive ion etching (CF$_4$ gas plasma), and e-beam metal evaporation (Cr 20 nm / Au 60 nm) were used to establish electrical edge-contact to few-layer graphite.[5]

**Magneto-transport measurement**

For the transistor devices, the bias voltage and gate voltage were applied using a homemade low-noise source meter and a Keithley 2400 SourceMeter, respectively. For the vertical tunnel junction devices, since resistance of the junction becomes comparable to resistance of graphite contacts, we performed four-terminal measurements with bias voltage applied using a homemade low-noise source and the voltage-drop measured with an Agilent 34410A. For both device types, the current was measured using a homemade low-noise current-voltage amplifier in combination with an Agilent 34410A. Magnetic field was applied in the plane of $NiPS_3$, and the relative angle between the applied magnetic field and the easy axis was set by aligning the sample chip on measurement stick of a cryostat at room temperature prior to the device cooldown. For the thicker samples (9 nm and 4 nm), the easy-axis direction (*a*-axis) was determined by polarization-resolved photoluminescence measurements.[5] The photoluminescence is extremely weak in tri- and bilayer samples, therefore for the ultrathin samples we determined the direction of the Néel vector using magnetotransport measurements. In particular, we performed a series of MR measurements with the magnetic field aligned along each of the three crystallographic edges of the flakes, revealing the unique axis along which the spin-flop is observed, which we then assigned to the easy axis.

**AUTHOR CONTRIBUTIONS**

C.Y. designed the experiments, fabricated the devices and performed the transport measurements. K.W. and T.T. grew the h-BN crystals. D.L. and A.F.M. supervised the project. C.Y., D.L., and A.F.M. analysed the data and wrote the manuscript.

**ACKNOWLEDGEMENTS**

The authors gratefully acknowledge Alexandre Ferreira for technical support. D.L. acknowledges the Swiss National Science Foundation (project PZ00P2_208760). A.F.M.

gratefully acknowledges financial support from Division II of the Swiss National Science Foundation, under project 200021_219424. K.W. and T.T. acknowledge support from the JSPS KAKENHI (Grant Numbers 21H05233 and 23H02052), the CREST (JPMJCR24A5), JST and World Premier International Research Center Initiative (WPI), MEXT, Japan.

**COMPETING INTERESTS**

The authors declare no competing interests.

**DATA AVAILABILITY**

The data that support the findings of this study are available from the corresponding author on reasonable request.

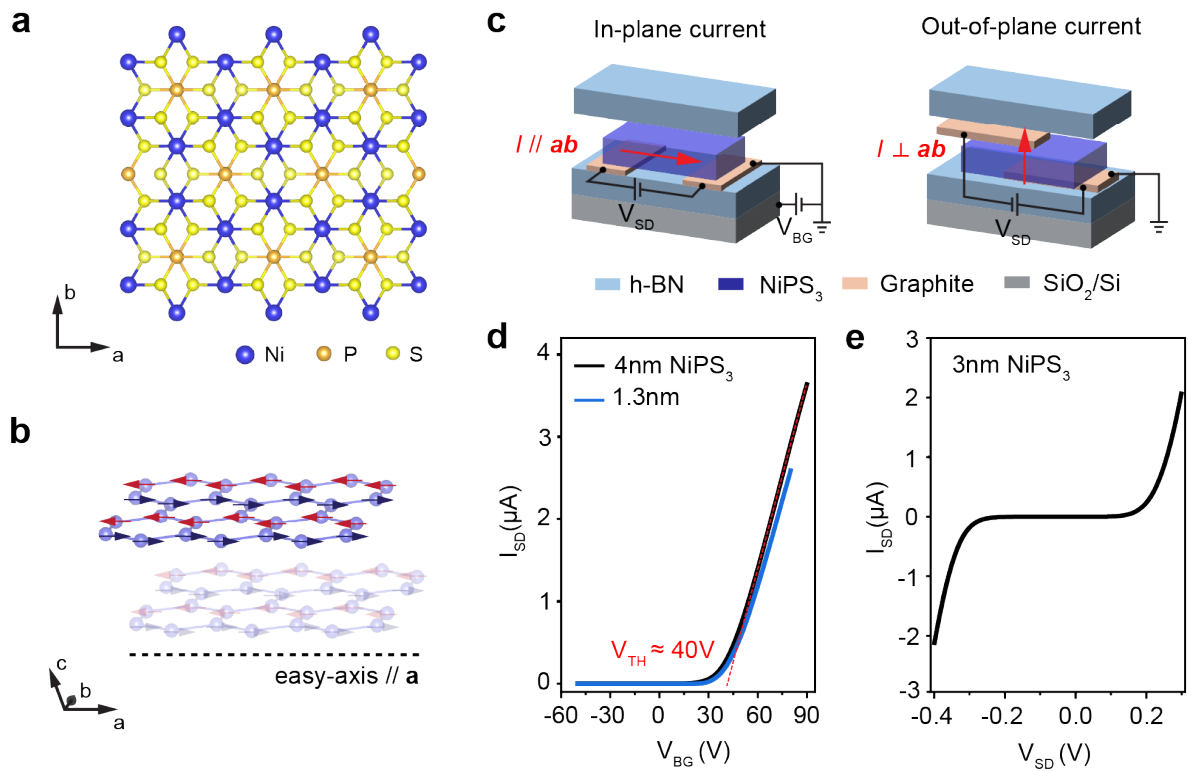

**Figure 1. NiPS₃ structure and device geometries. a.** Crystal structure of a single-layer NiPS$_3$. **b**. Monoclinic stacking and AFM order in a multilayer NiPS$_3$. Red and blue arrows indicate the collinear magnetic moments of the two sublattices, which are oriented along the crystallographic *a*-axis. **c**. Schematic of an FET (left) and a vertical tunneling device (right), where current flows in-plane and out-of-plane of NiPS$_3$, respectively. **d**. Transfer curves of 4-nm and 1.3-nm thick NiPS$_3$ FETs measured at a source-drain bias V$_{SD}$ = 4 V and a temperature T = 40 K. V$_{BG}$ and V$_{TH}$ denote the back-gate and the threshold voltages respectively. **e.** I-V curve of a 3-nm thick NiPS$_3$ tunneling barrier measured at T = 40 K.

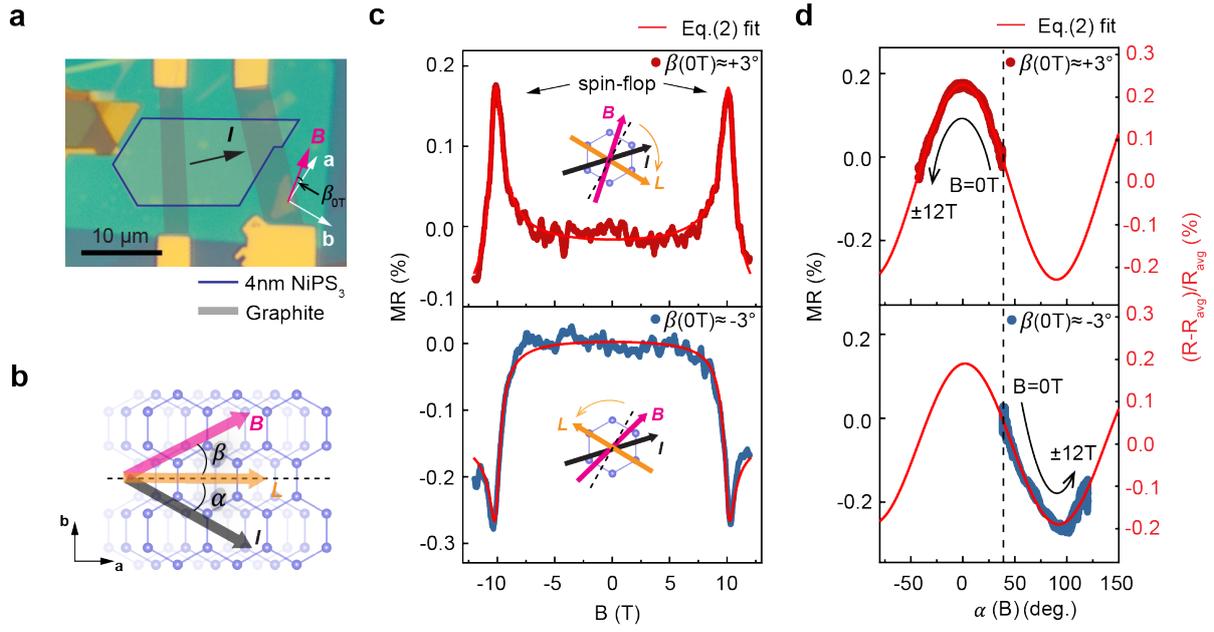

**Figure 2. Noncrystalline AMR in 4-nm NiPS$_3$. a.** Optical microscopy image of a 4-nm NiPS$_3$ FET (NiPS$_3$: blue contour line, graphite: gray-shaded area between gold contacts). $\beta(0T)$ denotes the angle between in-plane magnetic field **B** (pink arrow) and *a*-axis. The black arrow indicates the direction of electrical current (**I**), defined by the graphite strips. **b.** Schematic showing the relative orientation of **I** and **B** with respect to the crystal structure and the direction of the Néel vector (**L**). **c.** *MR* as function of magnetic field for $\beta(0T) \approx +3°$ (top panel) and $\beta(0T) \approx -3°$ (bottom panel) at T=40 K with V$_{SD}$= 4 V and gate offset $\Delta$V$_{BG}$ =V$_{BG}$-V$_{TH}$=40 V. **d.** *MR* (left) and normalized resistance (R-R$_{avg}$)/R$_{avg}$(%) (right) for $\beta_{0T} \approx +3°$ (top panel) and $\beta(0T) \approx -3°$ (bottom panel) as a function of $\alpha_B$, calculated using Eq. (1). The vertical black dashed line indicates the angle between the easy axis and **I**. The solid red line corresponds to the best fit using Eq. (2).

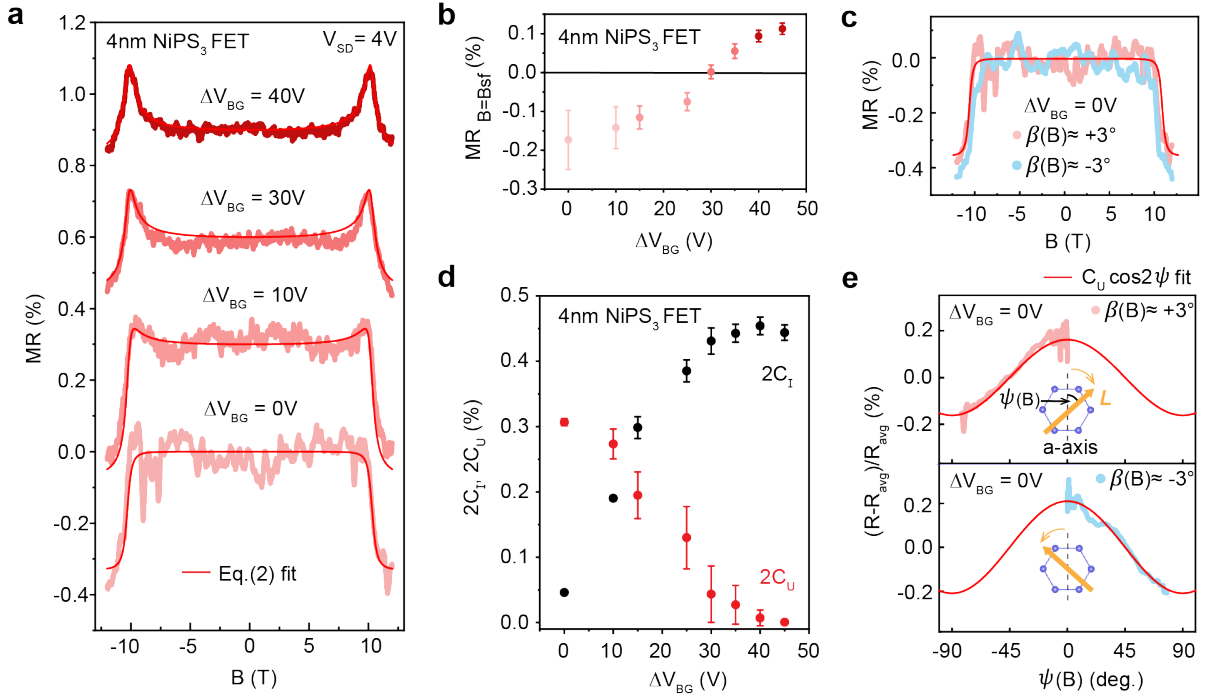

**Figure 3. Gate dependent AMR. a.** *MR* as a function of B measured at different gate offsets $\Delta V_{BG}$ for a 4-nm $NiPS_3$. Measurements are performed at fixed $V_{SD}$= 4 V and T=40 K. Curves are vertically offset for clarity. Red lines represent the best-fit curves based on the Eq. (2). **b.** *MR* at ($B=B_{sf}$) as a function of $\Delta V_{BG}$, showing the sign change as $\Delta V_{BG}$ approaches zero. The error bars represent the standard deviation of *MR*(%) calculated by averaging points in the vicinity of B = 0 T (from -3 T to 3 T). **c.** *MR* curves for two opposite field alignments measured at $V_{SD}$= 4 V and $\Delta V_{BG} \approx 0$ V. **d.** Noncrystalline ($C_I$) and crystalline ($C_U$) coefficients as a function of $\Delta V_{BG}$, extracted from the fitting. **e.** $(R-R_{avg})/R_{avg}$(%) for the two opposite field alignments ($\beta_{0T} \approx +3°$ for the top panel; $\beta_{0T} \approx -3°$ for the bottom panel) as a function of $\psi(B)$: an angle between the Néel vector and the easy axis, as illustrated in the inset schematics.

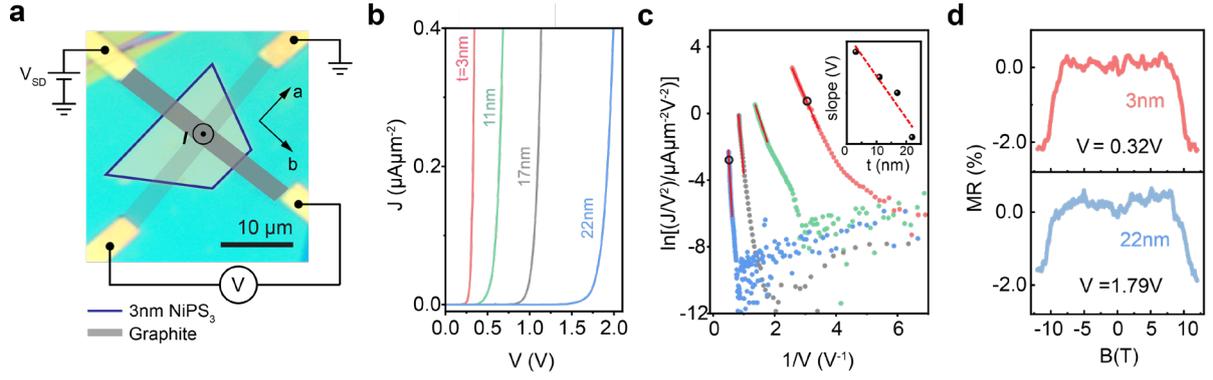

**Figure 4. Crystalline AMR in NiPS$_3$ tunnel junctions. a.** An optical microscopy image and the four-terminal measurement schematic of a vertical tunnel junction device based on 3-nm thick NiPS$_3$. **b.** Tunneling current density (J) measured at T=10 K as a function of V for different barrier thickness (t). **c.** Linear region (red line) between ln(J/V$^2$) and 1/V indicating Fowler−Nordheim tunneling at high bias. The inset shows the slope as a function of t, which follows a linear dependence (dashed red line). **d.** Tunneling MR as a function of B for 3-nm and 22-nm thick NiPS$_3$ with B aligned along the *a*-axis. Measurements are performed at T=10 K at bias voltages within the Fowler−Nordheim tunneling regime (indicated by open circles in panel **c**).

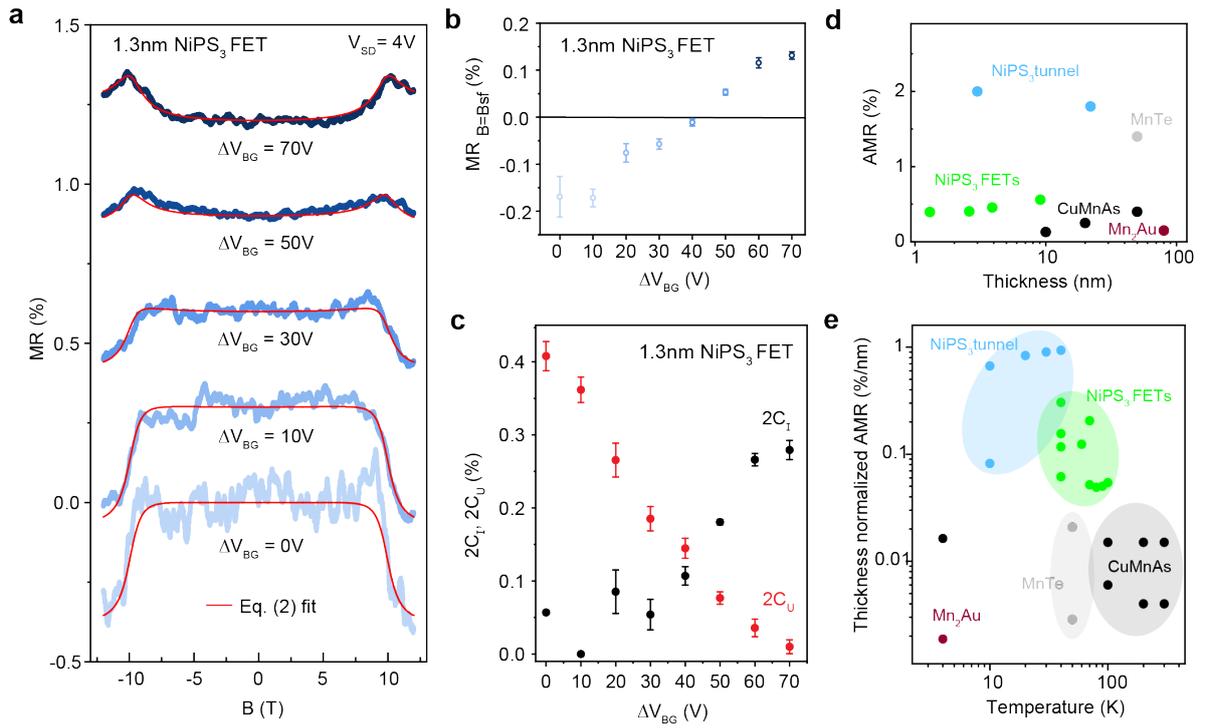

**Figure 5. AMR in ultrathin devices. a.** *MR* as a function of B measured at different gate offsets $\Delta V_{BG}$ for a bilayer (1.3-nm) NiPS$_3$. Measurements are performed at fixed $V_{SD}$= 4 V and T=40 K. Curves are vertically offset for clarity. Red lines represent the best-fit curves based on Eq. (2). **b.** *MR* at ($B=B_{sf}$) as a function of $\Delta V_{BG}$, showing the sign change as $\Delta V_{BG}$ approaches zero. The error bars represent the standard deviation of *MR*(%) calculated by averaging points in the vicinity of B = 0 T (from -3 T to 3 T). **c.** Noncrystalline ($C_I$) and crystalline ($C_U$) coefficients as a function of $\Delta V_{BG}$, extracted from the fitting. **d.** Comparison of the AMR magnitude (%) of NiPS$_3$ with other collinear AFM materials as a function of thickness; the largest low-temperature (T < 50 K) AMR values are shown. **e.** Thickness-normalized AMR magnitude (%) of NiPS$_3$ and other collinear AFM materials as a function of temperature. Data are taken from literature (ref[9] for CuMnAs, ref[23] for MnTe, ref[22] for Mn$_2$Au).

19